# Self-organization and magnetic domain microstructure of Fe nanowire arrays

N. Rougemaille and A.K. Schmid

Lawrence Berkeley National Laboratory, 1 Cyclotron Road MS/72, Berkeley, California 94720

**Abstract**

Starting from essentially flat nanometer-thick Fe films, epitaxially grown at room temperature on W(110) surfaces, we used carefully tuned annealing schedules to produce periodic arrays of nanoscale ferromagnetic wires. The structural transition from continuous films to nanowire arrays is accompanied with an in-plane 90° rotation of the spontaneous magnetization. Using spin-polarized low-energy electron microscopy to map the local magnetization directions while annealing, we studied the role of the dewetting mechanism on the self-organization and magnetization reorientation processes.

In ferromagnetic epitaxial thin films, the direction of spontaneous magnetization generally results from the interplay between magnetocrystalline, magnetoelastic, and shape anisotropies.[1] Since the balance between the corresponding magnetic anisotropy energies might be affected by a change in the film temperature or thickness, many systems exhibit reorientations of the magnetization as a function of these parameters.[2–5] Fe thin films grown epitaxially on W(110) surfaces are one interesting example of systems that show magnetization reorientation transitions. In Fe/W(110) films with thickness in the range of 1–2 atomic monolayers, magnetic properties are complex and, under some conditions, the easy axis of magnetization can be either in-plane or perpendicular to the film plane.[6–8] Thicker films are magnetized within the film plane. A transition can be observed near 10 nm thickness. Below this critical thickness, the easy axis of magnetization lies in the [1−10] direction, due to a strong surface anisotropy.[9,10] Above the critical thickness, the easy axis of magnetization switches to the in-plane [001] direction, as for bulk Fe.

Here we focus on a different type of magnetic transition, which can be observed when room-temperature-grown, nanometer-thick Fe/W(110) films are annealed for a few minutes between 700 and 1000 K. During annealing the films show an in-plane magnetization reorientation from the [1−10] to the [001] direction. In this case, the magnetization reorientation was shown to be induced by the formation of coalesced three-dimensional (3D) islands elongated along the [001] direction.[11–13] Different factors have been proposed as the driving mechanism for the rotation of the magnetic easy axis after annealing: reduction of the magnetoelastic anisotropy and contribution of the shape anisotropy,[13] or reduction of the surface anisotropy in the coalesced 3D islands.[11] In this work, we use spin-polarized low-energy electron microscopy (SPLEEM) to obtain spatially resolved magnetic information during the morphological transition. For our study, the main advantage of SPLEEM compared to other techniques is the rapid image acquisition (typically 1 frame/s for magnetic images and video rate for topographic images) that allows us to monitor in real time the dewetting process and its effect on the magnetization reorientation.[5,14]

Sample preparation and SPLEEM measurements were performed *in situ* under ultrahigh vacuum conditions. The Fe films were grown by molecular-beam epitaxy on a W(110) single crystal. The W substrate was cleaned by many cycles of flash heating to 2300 K under $3\times10^{-8}$ torr oxygen pressure, followed by a final temperature flash under UHV conditions. The quality of the surface was checked by Auger spectroscopy and low-energy electron diffraction. The Fe was deposited at room temperature in the microscope chamber and was annealed subsequently.

The general observation that the morphology of the islands depends on the annealing time and temperature was noted before.[13,15,16] We show here that these effects can be controlled and can be exploited to fabricate unusual self-organized structures. Our measurements show how regular arrays of Fe islands with extremely elongated shapes can be obtained and how island morphology depends on details of preparation conditions, including the substrate step structure. We compare results from two different W single crystals: one has a higher density of atomic surface steps and, in many regions of the surface, shows pronounced step bunching. The other crystal has a substantially lower density of steps and very large, atomically flat regions. Figure 1 shows topographic images of the two surfaces. We find that the combination of moderate annealing temperatures (between 600 and 650 K) and atomically flat substrates promotes the self-organization of remarkably regular periodic arrays of Fe wires.

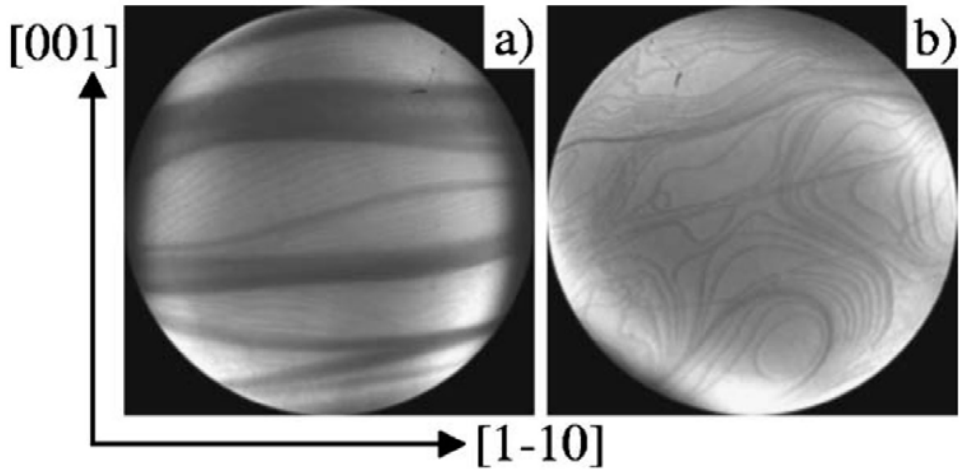

FIG. 1. Topographic images of the stepped (a) and flat (b) W substrates. Atomic steps appear as dark lines. The dark bands in image (a) are the regions of the surface showing pronounced step bunching. The field of view is 7 μm.

To illustrate the formation and magnetism of Fe/W(110) nanowire arrays, we focus on 14-monolayer-thick Fe films, which were grown on the two W crystals. Figures 2(a) and 2(e) show the surface topography of Fe films grown on the stepped and flat substrates, respectively, in an early stage of the dewetting process. On these images, the continuous films appear bright, while the dark regions are places where dewetting of the Fe films has occurred. In the dewetted regions, one pseudomorphic atomic Fe wetting layer remains on top of the W surface.[17] Due to the anisotropy of the W(110) substrate, voids formed in the Fe film during dewetting are preferentially elongated along the [001] direction, except on the step bunches where the high step density initiates the formation of compact-shaped voids along the step bands. With increasing annealing time, dewetted regions grow along the [001] direction, essentially forming very long, parallel grooves in the Fe film, accumulating the removed Fe along their sides. The dewetting process strongly affects the magnetization of these films, as shown in the corresponding SPLEEM images. We adjusted the spin polarization of the instrument such that in Figs. 2(b) and 2(f), bright (and dark) contrast corresponds to regions magnetized parallel (antiparallel) to the [1−10] direction, while in Figs. 2(c) and 2(g), bright and dark contrast corresponds to magnetization along the [001] direction. Note that no magnetic contrast (gray) is observed inside the grooves, since the pseudomorphic layer is magnetic only at cryogenic temperature.[6] The images shown in Figs. 2(d) and 2(h) highlight the dramatic influence of the substrate morphology: the Fe wires grown on the stepped substrate are all magnetized along the [001] direction [Fig. 2(d)], while Fe wires grown on a very flat substrate, under otherwise equivalent conditions, show a far more complex magnetic domain microstructure [Fig. 2(h)]. By observing the growth and magnetism of the wires from the start, we can attribute this difference to a subtle but important effect that takes place when the wires first nucleate. The onset of the dewetting process on flat W substrates takes place by nucleation and growth of rather elongated voids, as seen in Fig. 2(e).We observe strong magnetic contrast in Fig. 2(f) and no contrast in Fig. 2(g): this indicates that the Fe film maintains the original direction of magnetization, along [1−10], right up to the borders of the dewetted regions. The onset of dewetting is different on stepped W surfaces, where we observe the formation of chains of more compact voids in the Fe layer along pronounced substrate step bands, see Fig. 2(a). In the image shown in Fig. 2(c), magnetic contrast is visible in the regions adjacent to the voids. The contrast always appears black on the left side of the dewetting zones and white on the right side: this indicates in-plane canting of the magnetization of the Fe regions adjacent to the voids, presumably the canting is energetically favorable because the strength of stray fields in the voids is reduced. When the voids grow along the [001] direction during continued annealing, the canted magnetization in the nucleation region provides a bias to initiate reorientation of the magnetization along the wire direction. As a result, Fe/W(110) wires formed on stepped crystals normally have a single-domain magnetic structure, and wire length is limited by the distance between two step bands [Fig. 2(d)]. Fe/W(110) wires grown on flat substrates can be much longer (many-micrometers length is common) and we can form wires in which magnetization remains in the [1−10] direction, perpendicular to the growth direction of the wires. With further annealing, the width of these wires continues to shrink and, as a result, we find that at some point (typically when wire width decreases below 150 nm) the magnetization of the wires flips into the [001] direction. It is likely that increasing shape anisotropy, as a result of the increasing aspect ratio, is a factor in this reorientation of the magnetization. It is interesting to note that the process of this magnetization reorientation is often accompanied with the transient formation of a multidomain magnetic structure. Figure 3 shows an example where we used the SPLEEM to map these domain structures. A topographic image of a wire is shown in (a) and bright/dark contrast in three magnetic images shows the components of the magnetization along the [1−10] (b), [001] (c), and [101] (d) directions. On the upper and lower parts of the wire, the magnetization already flipped to the [001] direction, while the middle region of the wire exhibits flux-closure domains, as can be expected in wires having a strong in-plane uniaxial magnetocrystalline anisotropy, perpendicular to the wire axis.[18] The directions of magnetization in these domains, which we deduced from Figs. 3(b)–3(d), are sketched in Fig. 3(e). Contrary to what is observed in Fe films grown on stepped W(110) substrates, the magnetization reorientation in films grown on flat surfaces involves the nucleation of domains aligned with the growth direction of the wires, resulting from the competition between the magnetocrystalline anisotropy of the Fe/W(110) layer and the shape anisotropy of the elongated Fe wires.

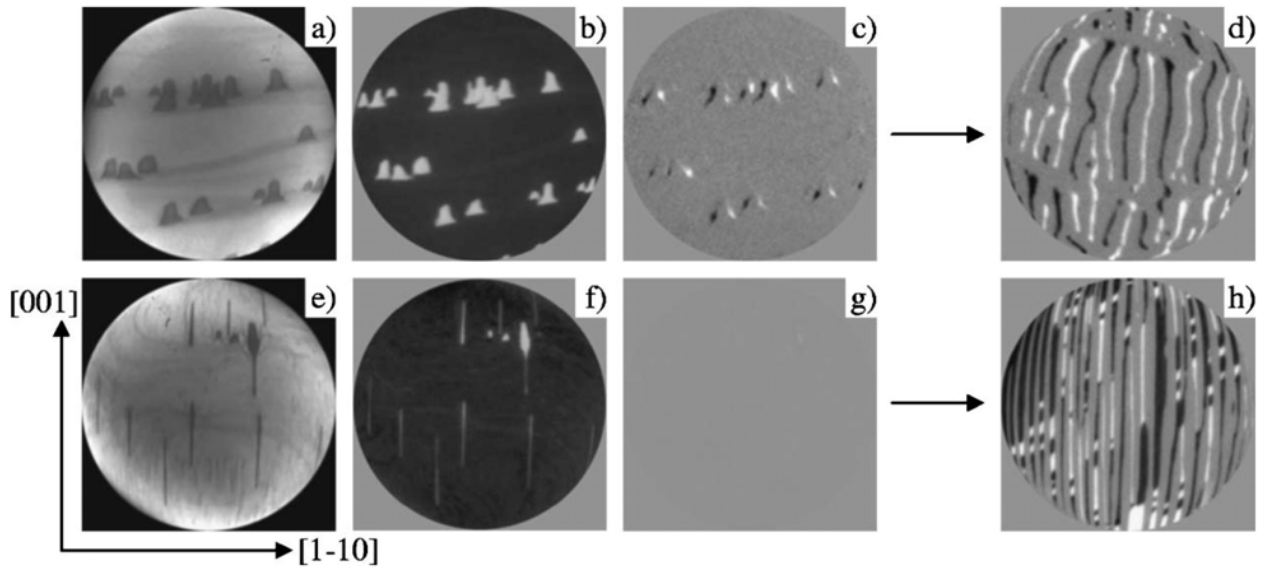

FIG. 2. Influence of the dewetting process on the magnetization reorientation. The field of view is 7 μm. Dewetted regions appear dark in the two topographic images [(a) and (e)]. The corresponding magnetic images for a spin polarization along the [1−10] [(b) and (f)] and [001] [(c) and (g)] directions show how the shape of the dewetted regions affects the magnetization distribution of the continuous film. Depending on the substrate morphology, complete annealing leads to the formation of single-domain Fe wires magnetized along the [001] direction (d), or Fe wires exhibiting more complex magnetic domain microstructure. See text for details.

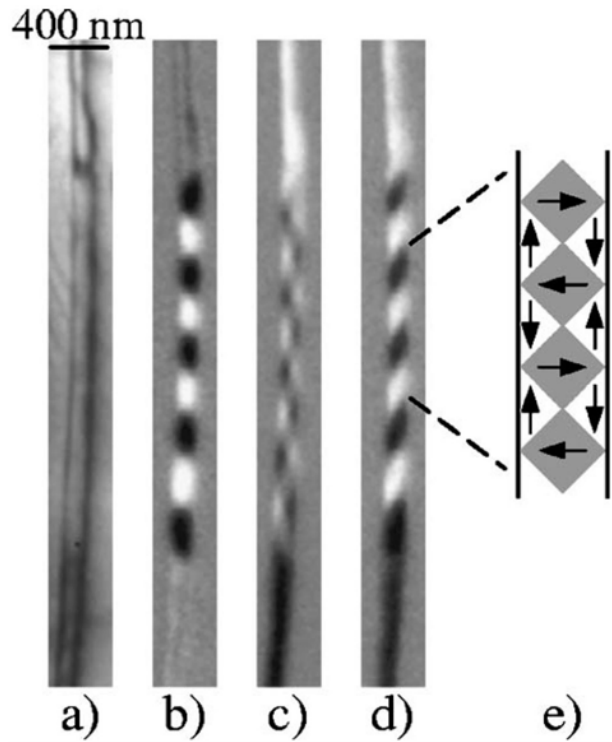

FIG. 3. (a) Topographic image (0.42×4.34 μm) of a wire obtained after annealing a 14-monolayer-thick flat Fe/W(110) film at 600 K. Corresponding magnetic images for an incident spin polarization along the [1−10] (b), [001] (c), and [101] (d) directions. (e) Sketch of the magnetization distribution in a wire resulting from the competition between surface and shape anisotropies.

Using spin-polarized low-energy electron microscopy, we studied the role of the dewetting process on the magnetization reorientation mechanisms observed in annealed, nanometer-thick Fe/W(110) films. At annealing temperatures near 650 K, flat W surfaces favor a dewetting process involving the nucleation of well-defined elongated voids, which do not promote canting of the magnetization of the Fe film. In contrast, on stepped substrate surfaces we observe the formation of more compact voids. The compact voids promote in-plane canting of the magnetization in adjacent regions of the Fe layer. As the voids grow, the regions with canted magnetization provide a bias that promotes the formation of single-domain Fe wires magnetized along their growth direction. Our findings highlight how the relative importance of different magnetic anisotropy energies contributing to this magnetic reorientation can be influenced by controlling the dewetting of the Fe film.

This work was supported by the U.S. Department of Energy under Contract Nos. DE-AC02-05CH11231 and DE-AC04-94AL8500, and by the Délégation Générale pour l'Armement under Contract No. 9860830051.

[1] *Ultrathin Magnetic Structures I*, edited by J. A. C. Bland and B. Heinrich (Springer, Berlin, Heidelberg, 1994).

[2] R. Allenspach, J. Magn. Magn. Mater. **129**, 160 (1994), and references therein.

[3] D. P. Pappas, K.-P. Kämper, and H. Hopster, Phys. Rev. Lett. **64**, 3179 (1990).

[4] Z. Q. Qiu, J. Pearson, and S. D. Bader, Phys. Rev. Lett. **70**, 1006 (1993).

[5] R. Ramchal, A. K. Schmid, M. Farle, and H. Poppa, Phys. Rev. B **69**, 214401 (2004).

[6] H. J. Elmers, J. Hauschild, H. Höche, U. Gradmann, H. Bethge, D. Heuer, and U. Köhler, Phys. Rev. Lett. **73**, 898 (1994).

[7] H. J. Elmers, J. Hauschild, H. Fritzsche, G. Liu, U. Gradmann, and U. Köhler, Phys. Rev. Lett. **75**, 2031 (1995).

[8] O. Pietzsch, A. Kubetzka, M. Bode, and R. Wiesendanger, Phys. Rev. Lett. **84**, 5212 (2000).

[9] U. Gradmann, J. Korecki, and G. Waller, Appl. Phys. A: Solids Surf. **39**, 101 (1986).

[10] R. Kurzawa, K.-P. Kämper, W. Schmitt, and G. Güntherodt, Solid State Commun. **60**, 777 (1986).

[11]D. Sander, R. Skomski, A. Enders, C. Schmidthals, D. Reuter, and J. Kirschner, J. Phys. D **31**, 663 (1998).

[12]O. Fruchart, M. Eleoui, J. Vogel, P. O. Jubert, A. Locatelli, and A. Ballestrazzi, Appl. Phys. Lett. **84**, 1335 (2004).

[13]L. Lu, J. Bansmann, and K. H. Meiwes-Broer, J. Phys.: Condens. Matter **10**, 2873 (1998).

[14]E. Bauer, T. Duden, and R. Zdyb, J. Phys. D **35**, 2327 (2002).

[15]V. Senz, R. Röhlsberger, J. Bansmann, O. Leupold, and K.-H. Meiwes-Broer, New J. Phys. **5**, 47 (2003).

[16]M. Bode, R. Pascal, and R. Wiesendanger, J. Vac. Sci. Technol. A **15**, 1285 (1997).

[17]U. Gradmann and G. Waller, Surf. Sci. **116**, 539 (1982).

[18]C. L. Dennis, R. P. Borges, L. D. Buda, U. Ebels, J. F. Gregg, M. Hehn, E. Jouguelet, K. Ounadjela, I. Petej, I. L. Prejbeanu, and M. J. Thornton, J. Phys.: Condens. Matter **14**, R1175 (2002), and references therein.